\documentclass[pra,twocolumn,showpacs,groupedaddress,superscriptaddress,aps,10pt]{revtex4-1}
\usepackage{bm,graphicx,amsmath}
\usepackage{placeins}
\usepackage{amssymb}
\usepackage{amsmath}
\usepackage{epsfig}
\usepackage{amssymb}
\usepackage{color}
\usepackage[colorlinks,linkcolor=blue,citecolor=blue]{hyperref}
\usepackage{subfigure}

\begin{document}

\title{Free Space Optical Polarization De-multiplexing and Multiplexing \\ by means of Conical Refraction}
\date{\today}



\author{Alex Turpin}
\affiliation{Departament de F\'{\i}sica, Universitat Aut\`{o}noma de Barcelona, E-08193 Bellaterra, Spain}
\author{Yurii Loiko}
\affiliation{Departament de F\'{\i}sica, Universitat Aut\`{o}noma de Barcelona, E-08193 Bellaterra, Spain}
\author{Todor K. Kalkandjiev}
\affiliation{Departament de F\'{\i}sica, Universitat Aut\`{o}noma de Barcelona, E-08193 Bellaterra, Spain}
\affiliation{Conerefringent Optics SL, Avda Cubelles 28, Vilanova i la Geltr\'u, E-08800, Spain}
\author{Jordi Mompart}
\affiliation{Departament de F\'{\i}sica, Universitat Aut\`{o}noma de Barcelona, E-08193 Bellaterra, Spain}

\begin{abstract}Polarization de-multiplexing and multiplexing by means of conical refraction is proposed to increase the channel capacity for free space optical communication applications. The proposed technique is based on the forward-backward optical transform occurring when a light beam propagates consecutively along the optic axes of two identical biaxial crystals with opposite orientations of their conical refraction characteristic vectors. We present experimental proof of usefulness of the conical refraction de-multiplexing and multiplexing technique by increasing in one order of magnitude the channel capacity at optical frequencies in a propagation distance of $4\,$m. 
\end{abstract}

\pacs{ocis: 060.2605, 060.4230, 160.1190.}

\maketitle


In optical communications, different properties of a light field, such as its intensity, wavelength, polarization, and orbital angular momentum (OAM), can be used to provide optical channels to efficiently transmit the information. Thus, for example, the capacity of a communication channel can be substantially increased if one multiplexes different wavelengths of various input optical carrier signals into a single channel by using the Wavelength Division Multiplexing (WDM) technique \cite{mik2000}. For a monochromatic laser beam, Laguerre--Gauss light beams carrying orbital angular momentum (OAM) in the helicity of their phase fronts have been proposed \cite{all1992} as a basis of carrier signals allowing, in principle, for an arbitrary increase of the channel capacity \cite{gib2004,slu2011,gat2011,wan2011}. However, there are practical drawbacks that restrict the range of applicability of the OAM encoding technique \cite{jac2008,fra2004} such as the large divergence of high order OAM modes, which prevent their use for free space optical communications (FSOC) at long distances. Alternatively, one could also use the polarization degree of freedom of a light beam as a carrier basis of signals for FSOC links. In this case, nevertheless, the use of a polarization beam splitter allows, at most, to double the FSOC channel capacity. 
In this paper, we report a novel method to de-multiplex and multiplex a monochromatic input light beam into, in principle, an arbitrary large number of polarization states by means of the conical refraction phenomenon \cite{bel1978,bel1999,ber2007,kal2008,ber2010,abd2011,crlas2011,tur2011}. 

\begin{figure}
\centerline{\includegraphics[width=0.90 \columnwidth]{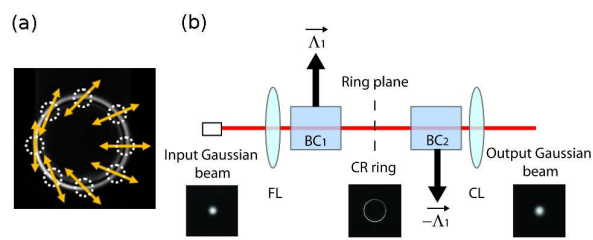}}
\caption{(color online) (a) Transverse intensity pattern registered at the focal plane for a circularly polarized input Gaussian beam after passing through a $28\,\rm{mm}$ long KGd(WO$_4$)$_2$ biaxial crystal. Yellow double arrows illustrate the polarization distribution along the ring for vertical orientation of $\vec{\Lambda}$. Dashed circle represent linearly polarized beams. (b) Degenerated 2-cascade conical refraction configuration yielding a forward-backward transform of a circularly polarized input Gaussian beam. The ring plane coincides with the focal plane of the first lens. FL: focusing lens. CL: collimating lens. BC$_1$: first biaxial crystal. BC$_2$: second biaxial crystal.}
\label{fig1}
\end{figure}

In conical refraction, when a circularly polarized collimated light beam passes along the optic axis of a biaxial crystal it refracts conically inside the crystal and emerges as a collimated hollow cylinder whose transverse profile is a light ring. This light ring is laterally shifted being both the direction of the displacement as well as the ring radius given by the so-called characteristic $\vec{\Lambda}$ vector of the biaxial crystal \cite{kal2008}. Each point of the light ring is linearly polarized with the polarization plane rotating continuously along the ring in such a way that every two opposite points of the ring have orthogonal polarizations, see Fig.~\ref{fig1}(a). This polarization distribution depends only on the orientation of $\vec{\Lambda}$.

When two identical biaxial crystals are placed with aligned optic axes and opposite orientations of their $\vec{\Lambda}$ vectors, see Fig.~\ref{fig1}(b), an input Gaussian beam is transformed, after propagating through the first crystal, into a light ring that, after passing through the second crystal, gives back the input Gaussian beam \cite{tur2011}. This particular arrangement, which allows optical forward-backward transform, will be called in what follows as the degenerate 2-cascade conical refraction configuration.
 
We make use of this forward-backward transform of conical refraction to propose a novel method to de-multiplex and multiplex a monochromatic light beam into a large number of linearly polarized states as it is schematically shown in Fig.~2. The first crystal de-multiplexes the input beam into an infinite number of linearly polarized beams placed along a ring (see Fig.~\ref{fig1}(a) where each dashed circle represents a linearly polarized beam). Each of these beams constitutes an information channel (note that the channels are polarization channels) that can be individually selected and modulated in amplitude. Later on, the second biaxial crystal multiplexes all the channels into one beam that propagates in free space. Finally, a third biaxial crystal can be used to decode the transmitted signal at the receiver stage. In what follows we describe the conical refraction de-multiplexing and multiplexing protocol in terms of the standard elements that form a free space optical telecommunications system: the transmitter, the free space propagation, and the receiver. The mutual alignment of the optic axes of the biaxial crystals should be maintained with precision within 50 $\mu$rad 
to make it work properly by means of the CR effect.

\begin{figure}[ht!]
\centerline{\includegraphics[width=0.90 \columnwidth]{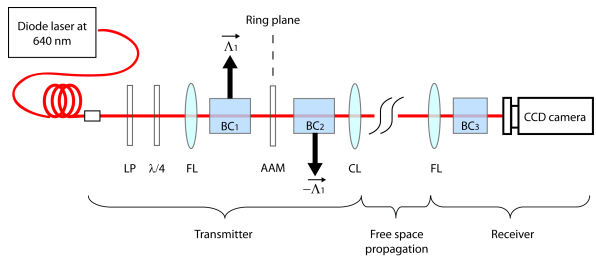}}
\caption{Sketch of the polarization de-multiplexing and multiplexing experimental set-up based on conical refraction being used to increase the channel capacity of a FSOC system. The FSOC system is formed by the transmitter with its two identical biaxial crystals presenting opposite $\vec{\Lambda}$ vectors, a free space propagation distance of $4\,$m, and the receiver with the third biaxial crystal. LP: linear polarizer. $\lambda/4$: quarter wave plate. AAM: angular amplitude mask.}
\label{fig2}
\end{figure}

The transmitter consists of an input monochromatic light beam, two biaxial crystals in a degenerate 2-cascade configuration, an angular amplitude mask, and the lenses to focus and collimate the beam. As input beam, we take a collimated linearly polarized Gaussian beam with $w_0=1\,\rm{mm}$ beam waist obtained from a $640\,\rm{nm}$ diode laser coupled to a monomode fiber. A linear polarizer and a quarter wave plate are placed to ensure a perfect circularly polarized Gaussian beam at the entrance of the first crystal. Note that the experiment could also be performed with a linearly polarized input beam but with the inconvenience of producing a crescent intensity pattern instead of a complete ring and, therefore, the polarization channels would possess different amplitudes. The degenerated 2-cascade scheme is prepared with two identical KGd(WO$_4$)$_2$ biaxial crystals ($<100\,\rm{nm}$ of difference) yielding a light ring after the first crystal of $872\,\rm{\mu m}$ ring radius. The polished entrance surfaces of the two biaxial crystals (cross-section $6\times4\,\rm{mm}$$^2$) have parallelism with less than 10 arc seconds and they are perpendicular to one of the two optic crystal axes within $1.5\,\rm{mrad}$ misalignment angle. To focus and collimate the beam we use lenses with $200\,\rm{mm}$ focal length. To select the polarization channels at the light ring we use angular amplitude masks forming a star burst-like pattern with $n$ (up to $12$) opened circular sectors. The amplitude masks actually allow passing only some parts of the ring, thus we are indeed selecting the communication channels. Encoding the information into the different channels could be performed by time varying the transmission coefficient for each sector of the mask using, for instance, a spatial light modulator.

In our experiments, the free space propagation distance is $4\,\rm{m}$. We have measured that the multiplexed beam has a divergence similar to the initial Gaussian beam and, therefore, we expect that our protocol could operate for the same distances as other FSOC systems do it with Gaussian beams. Results on the extension of our protocol to larger propagation distances will be presented elsewhere.

\begin{figure}[ht!]
\centerline{\includegraphics[width=0.90 \columnwidth]{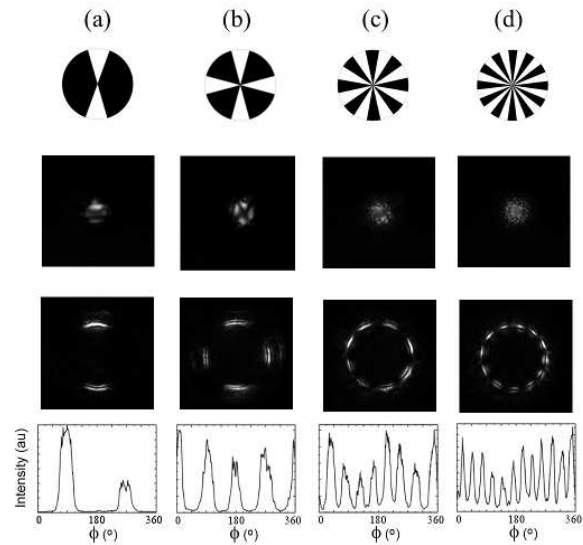}}
\caption{Experimental transverse intensity patterns (third row) and the corresponding integrated azimuthal intensities (fourth row) obtained by the receiver for multiplexer masks (first row) with $2$ (a), $4$ (b), $8$ (c), and $12$ (d) opened sectors. The second row shows the multiplexed beams at the entrance of the receiver.}
\label{fig3}
\end{figure}

\begin{figure}[ht!]
\centerline{\includegraphics[width=0.90 \columnwidth]{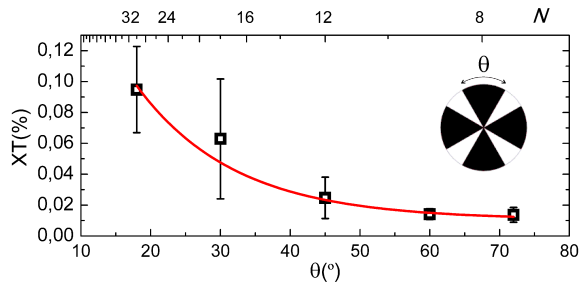}}
\caption{Crosstalk (XT) between adjacent channels vs. the closure angle $\theta$ (bottom axis) and vs. the number of channels $N$ (top axis) of the masks used. Red solid curve represents exponential fitting to the experimental data. Uncertainty of the $\theta$ angle measurement was below $1^o$.}
\label{fig4}
\end{figure}

The receiver itself consists of an objective of $50\,\rm{mm}$ focal length, a $12\,\rm{mm}$ long KGd(WO$_4$)$_2$ biaxial crystal, and a CCD camera. This biaxial crystal de-multiplexes (final patterns shown in the third row of Fig.~\ref{fig3}) the free space propagated beam (transversal patterns shown in the second row of Fig.~\ref{fig3}), performing conical refraction and recovering the sectors that were modulated by the angular amplitude masks (first row of Fig.~\ref{fig3}) at the transmitter. As it can be observed in the third row of Fig.~\ref{fig3}, we are able to independently modulate up to 12 sectors, which constitutes an increase in one order of magnitude of the channel capacity of the FSOC link. Last row in Fig.~\ref{fig3} shows the intensity variation along the azimuthal direction of the corresponding de-multiplexed patterns from the third row of Fig.~\ref{fig3}. The intensity peaks of the received channels are perfectly distinguishable with respect to the background. Additionally, one can also note that there is no crosstalk between neighbor channels, since there appear as number of peaks as number of channels selected at the transmitter.
\\
Crosstalk (XT) between the channels is one of the main limiting factors for real applications. The main contribution into the XT between adjacent channels in our system comes from light diffraction on the mask domain boundaries. To characterize the XT, we have investigated the influence of the closure angle of the masks, i.e. the azimuthal angle separating neighbor open sectors (see $\theta$ in the inset of Fig.~\ref{fig4}), and the number of channels over it by measuring the residual intensity at the center of the closed sector.  For the latter case, the open and closed sectors in the mask have the same azimuthal angular width and we measure the XT at the closed sectors. The results for XT, i.e. residual intensity related to the intensity maximum, are presented in Fig.~\ref{fig4}. The data reveal that, as it can be expected, the smaller the number of channels the smaller the XT. Moreover, the thinner the open sectors, which corresponds to larger closure angle $\theta$, the smaller the XT too. Red solid curve gives exponential fitting to the experimental data that show the XT decay as $\theta$ increases and $N$ decreases. For the 12 channels case shown in Fig.~\ref{fig3}, the average XT is less than $3\%$. Finally, we would like also to note that misalignment in crystals' rotation around the beam propagation direction leads to light polarization XT between any opposite points at the CR ring. However, in our system it is well controlled below $10^{-6}$.

In summary, we have proposed a novel technique to de-multiplex and multiplex a monochromatic light beam into a finite and, in the ideal case, an arbitrary number of linearly polarized states. The technique is based on the forward-backward transform produced by two biaxial crystals under conditions of conical refraction. We have demonstrated an increase of one order of magnitude in the channel capacity for FSOC of a monochromatic input Gaussian beam at $640\,\rm{nm}$ for a $4\,\rm{m}$ propagation distance with cross-talk being below $3\%$. In addition, we have investigated the XT with respect to the azimuthal angle of the closed sectors and the number of sectors of the masks used. The obtained results suggest that by simply optimizing the channel selecting mechanism, i.e. the thickness of the open and closed sectors of the masks, one could increase the channel capacity or decrease the XT for a fixed number of channels. Details on such optimization will be reported elsewhere.

As an encouragement for future investigations on the technique proposed in this paper, we would like to note that by selecting appropriate biaxial crystals it would be interesting to extend this novel method to other wavelengths in the optical and telecommunication bands at which the crystals are transparent and to combine it with the WDM technique. Finally, it would be very promising to look for new quantum cryptography protocols by extending the technique to the single-photon case.

The authors gratefully acknowledge financial support through Spanish MICINN contracts CSD2006-00019, FIS2008-02425, FIS2010-10004-E, and FIS2011-23719, and the Catalan Government contract SGR2009-00347. A. T. acknowledges financial support through grant AP2010-2310 from the MICINN.

\end{document}